\documentclass{ws-mpla}

\newcommand{\arcsec}{\hbox{$^{\prime\prime}$}}

\begin{document}

\markboth{Christian Knigge}
{Hunting for the Products of Stellar Collisions and Near Misses}

\catchline{}{}{}{}{}
%
%%%%%%%%%%%%%%%%%%%%%%%%%%%%%%%%%%%%%%%%%%%%%%%%%%%%%%%%%%%%%%%%%%%%

\title{Far-Ultraviolet Surveys of Globular Clusters:\\Hunting for the 
Products of Stellar Collisions and Near Misses\footnote{
Based on observations with the NASA/ESA Hubble Space Telescope (HST),
obtained at the Space Telescope Science Institute (STScI), which is
operated by the Association of Universities for Research in
Astronomy, Inc., under NASA contract NAS 5-26555}}

\author{\footnotesize CHRISTIAN KNIGGE}

\address{School of Physics and Astronomy, University of Southampton\\
Southampton SO17 1BJ, United Kingdom\\
christian@astro.soton.ac.uk}

\maketitle

\pub{Received (Day Month Year)}{Revised (Day Month Year)}

\begin{abstract}

Globular clusters are gravitationally bound stellar systems containing
on the order of $10^5$ stars. Due to the high stellar densities in the
cores of these cluster, close encounters and even physical collisions
between stars are inevitable. These dynamical interactions can produce
exotic types of single and binary stars that are extremely rare in the
galactic field, but which may be important to the dynamical evolution
of their host clusters. A common feature of these dynamically-formed
stellar populations is that many of their members are relatively hot,
and thus bright in the far-ultraviolet (FUV) waveband. In this short
review, I describe how space-based FUV observations are being used to
find and study these populations.

\end{abstract}

\ccode{PACS Nos.: include PACS Nos.}

\section{Introduction}

\subsection{Dynamically-Formed Stellar Populations}

Globular clusters (GC) are co-eval, gravitationally bound stellar
systems containing between $10^4$~and $10^6$~stars. Stellar densities
in the cores of GCs are often extremely high and can reach
$10^6$~stars per cubic parsec. Close encounters between cluster
members are therefore relatively common, which makes GC cores
fantastic stellar crash test laboratories.  

The most extreme type of close encounter is a direct physical
collision between two stars. It has been estimated that up to 40\%~of
the stars in the cores of the densest GCs may have suffered such a
collision during their lifetime.\cite{hills} One 
possible outcome of such a collision is a rejuvenated main sequence
star that is more massive (and therefore hotter) than either of its
two progenitors. Such collision products may be found along an
extension of the cluster main sequence (MS), but bluewards of the MS
turn-off. They are thus usually referred to as ``blue stragglers''
(BSs) and are observed in essentially all galactic GCs.\cite{shara}

Near misses are even more common than direct collision and are thought
to produce numerous close binary systems (CBs). The tightest and most
interesting CBs are those containing at least one compact object,
i.e. a white dwarf (WD) or a neutron star (NS). The most important 
and extreme compact binaries are those that are so tight that the
(usually MS) companion of the compact object cannot fit into its
``Roche lobe'', i.e. its part of the dumb-bell-shaped critical
gravitational potential of the binary. Matter then flows from the
companion to the compact object via the inner Lagrangian point, in a process called
Roche lobe overflow. The material lost by the mass donor is ultimately
accreted by the compact object, but in doing so releases significant
amounts of gravitational potential energy. Most of this emerges as
radiation in the X-ray and/or far-ultraviolet (FUV) wavebands. If the
accreting compact object is a NS, the system is called a low-mass
X-ray binary (LMXB); if it is a WD, the system is called a cataclysmic
variable (CV). For more information on these types of accreting
binaries, see Ref.~\refcite{lewin} (for LMXBs) and Ref.~\refcite{warner}
(for CVs).

There are several ways to create accreting compact binaries via 
dynamical encounters in GC cores. The first and most famous is tidal
capture. This is a 2-body interaction which relies on the high stellar
densities in the cluster core to bring a NS or WD very close to a MS
star. The MS star 
then experiences tidal distortions which dissipate orbital energy, and
can lead to capture and binary formation.\cite{fabian} 
However, compact binaries can also be formed via processes
involving existing binaries, i.e. 3- and 4-body
interactions. For example, in a close encounter between a low-mass
(e.g. MS/MS) binary system and a high mass (e.g. NS) single star, the
most likely outcome is ejection of the lowest mass participant and
formation of a NS/MS binary system.\cite{sigurd}

Tidal capture was originally proposed in order to account for the
observed 100-fold overabundance of LMXBs in GCs (relative to the
galactic field).\cite{fabian} However, this and other dynamical
formation processes might be expected to create a similar
overabundance of CVs. If so, the number of CVs should far outstrip the
number of LMXBs, since WDs are far more  
common than NSs. For example, the authors of Ref.~\refcite{distefano} predict
that roughly 200 CVs produced by tidal capture should
exist today in 47~Tucanae, and Ref.~\refcite{davies} notes that there could be
an additional 100 or so formed via 3-body interactions (as well as
perhaps another 300 systems in the cluster outskirts that are
descended directly from primordial binaries). However, finding these
systems proved to be a serious observational challenge,
and only recent surveys with modern, space-based X-ray
(Refs.~\refcite{grindlay1,grindlay2,pooley1,pooley2}; see
Refs.~\refcite{edmonds1,edmonds2} for the most detailed follow-up to
one of these X-ray surveys to date.) and far-ultraviolet detectors
(Refs.~\refcite{knigge1,knigge2}) are beginning to reveal a sizeable
population of CVs in GCs. Whether this population is large enough to
match the theoretical predictions remains to be seen (see also 
Section~\ref{state_image}).

\subsection{Relationship to Cluster Dynamics and Evolution}

Once they are formed, the dynamically-created stellar populations can 
actually themselves become key players in controlling further cluster
evolution. In other words, the ``interplay'' between cluster dynamics
and stellar evolution is very much a two way street. CBs are
particularly important in this context, since
the binding energy of even a single very close binary can rival that 
of an entire modest-size globular cluster (GC). As a result, the dynamical
evolution of a GC can in principle be driven by just a few
close binaries among its population. In practice, these binaries drive
cluster evolution by transferring their orbital energy to passing
single stars. Close binaries thus ``harden'', while promoting
cluster expansion and evaporation on a time-scale somewhat longer than
a Hubble time (see, for example, Ref.~\refcite{hut}).

\subsection{The Advantages of Far-Ultraviolet Surveys}

All of the most interesting dynamically-formed stellar populations --
BSs, LMXBs and CVs -- share a key observational characteristic: their
spectral energy distributions are much bluer than those of ``normal''
cluster stars. This immediately implies that FUV imaging should be an
excellent way to find and study these populations. However, the FUV
waveband has historically been almost inaccessible to GC imaging
studies. The reason for this is simple: FUV imagers with the
sensitivity and spatial resolution required to locate faint FUV-excess 
objects in GC cores (and without significant red leaks) simply did not
exist.\footnote{It is important, however, to acknowledge the
pioneering FUV and NUV studies that were 
carried with earlier generation detectors on HST (including, but not
limited to, Refs.~\refcite{ferraro1,paresce3,demarchi1,demarchi2,burgarella,demarchi3,ferraro2})
and also with the {\it Ultraviolet Imaging Telescope}
(Refs.~\refcite{hill,oconnell}). These studies already demonstrated
the great advantages of the UV waveband and detected both BSs and CV candidates.}

The arrival of first the {\em Space Telescope Imaging Spectrograph}
(STIS) and now the {\em Advanced Camera for Surveys} (ACS) on the {\em
Hubble Space Telescope} (HST) has radically improved this situation.
Both of these instruments feature sensitive, solar-blind UV
detectors whose spatial resolution ($\simeq 0.05$\arcsec) is actually
{\em better} than that of most of HST's optical imagers. As a consequence, the
first truly deep, far-UV imaging studies of GC cores are finally being
carried out.\cite{brown,knigge1,knigge2}. In this short review, I will present
the first results that have come out of these studies in relation to
the dynamically-formed stellar populations and look forward to how
this field is likely to develop in the near future.

\section{The Current State of Play}	

Deep FUV observations have so far been published for only two
clusters: 47 Tuc (work by our own group;
Refs.~\refcite{knigge1,knigge2}) and NGC~2808
(Ref.~\refcite{brown}). However, Ref~\refcite{brown} was focused 
primarily 
on hot, subluminous horizontal branch stars, which are not obviously
affected by dynamical encounters.\footnote{It should be noted,
however, that the counterparts of these hot, subluminous horizontal
branch stars in the galactic field -- the so called sub-dwarf O and B
stars -- are known to have an extremely high binary
fraction.\cite{maxted} In the context of a GC environment, this might suggest
that cluster dynamics is important for these stars also (since
binaries can be formed dynamically and can easily interact with other
cluster members). However, Ref.~\refcite{brown} note that the stars they
studied are not concentrated towards the cluster core and thus are
probably the product of single star evolution. Nevertheless, the
origin of these stars and their possible relation to cluster dynamics
clearly deserves further study.} Thus only our own FUV work on 47~Tuc
has so far focused on the dynamically-formed stellar populations. As a
result, the discussion in this section is necessarily
limited to only one cluster. However, our FUV survey of 47~Tuc did
include both imaging and slitless spectroscopy components, so I will
present the results emerging from our application of these two methods
in turn.
  
\subsection{FUV Imaging: Blue Stragglers and Cataclysmic
Variables in the Core of 47 Tucanae} 
\label{state_image}

We had two reasons for taking the prototypical cluster 47~Tuc as our 
initial FUV target. First, its observational properties -- in
particular, the angular size of its core and the lack of unduly
FUV-bright blue horizontal branch stars -- turn out to be well-matched
to the requirements of a FUV survey.\footnote{The presence of blue
horizontal branch stars is in principle not a problem, but makes it
more difficult to find faint sources that may be hidden in bright PSF
wings. This issue was particularly important for the slitless
spectroscopy part of our program and was therefore relevant for our
target selection.} Second, most of the theoretical work on dynamically
formed stellar populations has used this cluster as a point of
reference, so there are some clear predictions for what we 
{\em should} see. 

In total, we obtained thirty orbits of STIS/HST observations of 47
Tuc, comprised of six epochs of five orbits each (HST program
GO-8219). In each epoch, we carried out FUV imaging and slitless
spectroscopy. Consecutive observing epochs were as closely spaced as a
few days and as widely separated as a year. Typical exposure times in
the FUV were 600~s. Our program is therefore sensitive to variability on
time-scales ranging from minutes to months. This is important, since
accreting binaries tend to be variable on all of these time-scales. 

All of our FUV observations (both imaging and spectroscopic) used the
FUV-MAMA detectors and were taken through the F25QTZ filter. The
purpose of this filter is to block geocoronal Ly$\alpha$, OI
1304~\AA~and OI] 1356~\AA~emission which would otherwise produce a high 
background across the detector (this is especially important for our
slitless spectroscopy). The effective bandpass with this instrumental
set-up is 1450~\AA~--~1800~\AA. The 1024$\times$1024 pixel FUV-MAMA detector
covers approximately 25\arcsec$\times$25\arcsec, at a spatial
resolution of about 0.043\arcsec~(FWHM). Our field of view (FoV) was
chosen to overlap with archival HST observations of 47~Tuc and
included the cluster center. For comparison, the core radius of 47~Tuc
is 23 \arcsec.\cite{howell}

Figure~1 shows a comparison of the F336W and FUV images of the
same 25\arcsec$\times$25\arcsec~field near the core of 47~Tuc. Note
that the optical image is vastly more crowded, because the majority of 
main-sequence stars, red giants and horizontal branch stars are too
cool to show up in the FUV exposure. This is perhaps {\em the} key
advantage of FUV over optical surveys.

\begin{figure}[thp]
\centerline{\psfig{file=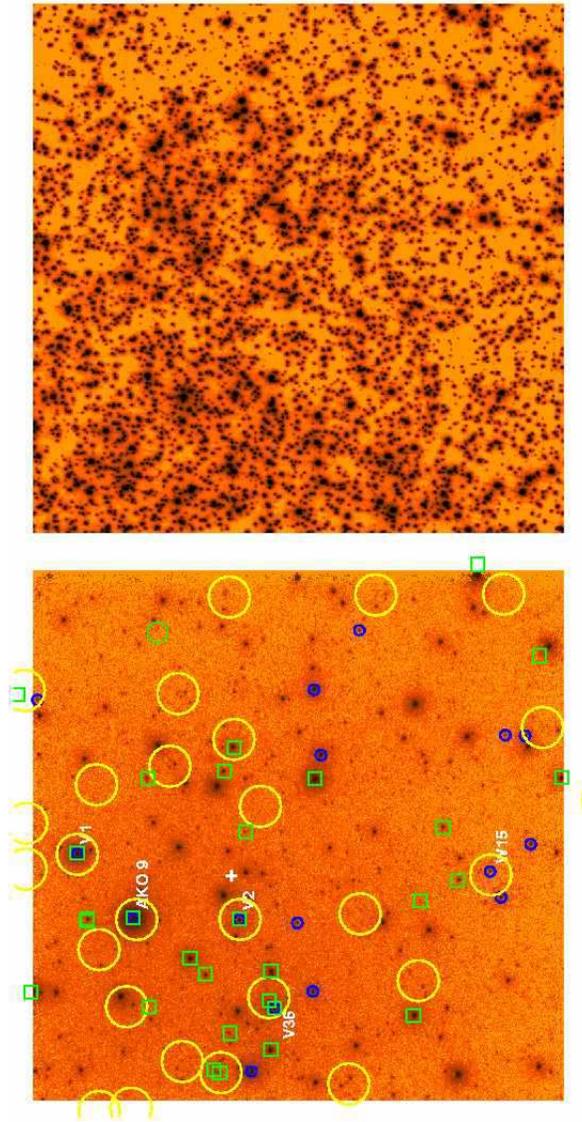,width=3.05in,angle=0}}
\vspace*{8pt}
\caption{{\em Bottom Panel:} The co-added FUV image of the core of
47~Tuc. The image is approximately 25\arcsec$\times$25\arcsec in size
and includes the cluster center (marked as a white cross; position
taken from Ref.~\protect\refcite{guha}). For comparison, 47~Tuc's core radius is
23\arcsec.\protect\cite{howell} The positions of known blue objects
from Ref.~\protect\refcite{geffert} (green squares), Chandra X-ray sources
from Ref.~\protect\refcite{grindlay1} (large yellow circles) and of the
objects in the ``CV zone'' (see text; small blue circles) are marked. The
four confirmed CVs within the FoV are labelled with their
most common designations. The
image is displayed on a logarithmic intensity scale and with limited
dynamic range so as to bring out some of the fainter FUV sources.
{\em Top Panel:} The co-added WFPC2/F336W image of the same
field. This image, too, is shown with a logarithmic intensity scale
and limited dynamic range. Figure reproduced from
Ref.~\protect\refcite{knigge1} (\protect\copyright~2002 The American Astronomical
Society.)}
\end{figure}

We have overlayed onto the FUV image the predicted positions of known
blue sources from Ref.~\refcite{geffert} and also the predicted positions
of the Chandra X-ray sources from Ref.~\refcite{grindlay1}
Almost all of the blue sources listed by Ref.~\refcite{geffert} have
certain or likely FUV counterparts. This was to be expected, but
is nevertheless 
important: it provides an external check on our astrometry and
supports our contention that FUV imaging is an excellent way of
finding interesting GC sources. We find fewer obvious FUV counterparts
to the X-ray sources found by Ref.~\refcite{grindlay1} within our
FoV. This is not so surprising, given that 
approximately 70\% of their sources are expected to be very faint FUV
sources (millisecond pulsars and X-ray active main sequence
binaries). A more interesting question is whether we detect those
Chandra sources that are definitely expected to be FUV
bright. There are four such sources within our FoV, all of
which were classified by Ref.~\refcite{grindlay1} as likely CVs. Three
of these are thought to be associated with previously known or
suspected CVs, namely AKO~9, V1 and V2; all three are also clearly detected in our
FUV images (and indeed confirmed as CVs by our FUV spectroscopy; see
Section~\ref{state_spec}). The fourth source -- 
denoted W15 in Ref.~\refcite{grindlay1} -- is a CV candidate that was not
known prior to the Chandra survey of 47~Tuc. It, too, has a likely
counterpart in our FUV images (see Figure~1).

Figure~2 shows the FUV-optical color-magnitude diagram (CMD) of
47~Tuc. Several distinct stellar populations are present in CMD, 
among them WDs, BSs, MSTO stars and, last but not least, CVs. In order
to aid in the interpretation of the CMD, we have also calculated and
plotted a set of theoretical tracks. Two types of dynamically-formed
populations can be spotted immediately. 

\begin{figure}[ht]
\centerline{\psfig{file=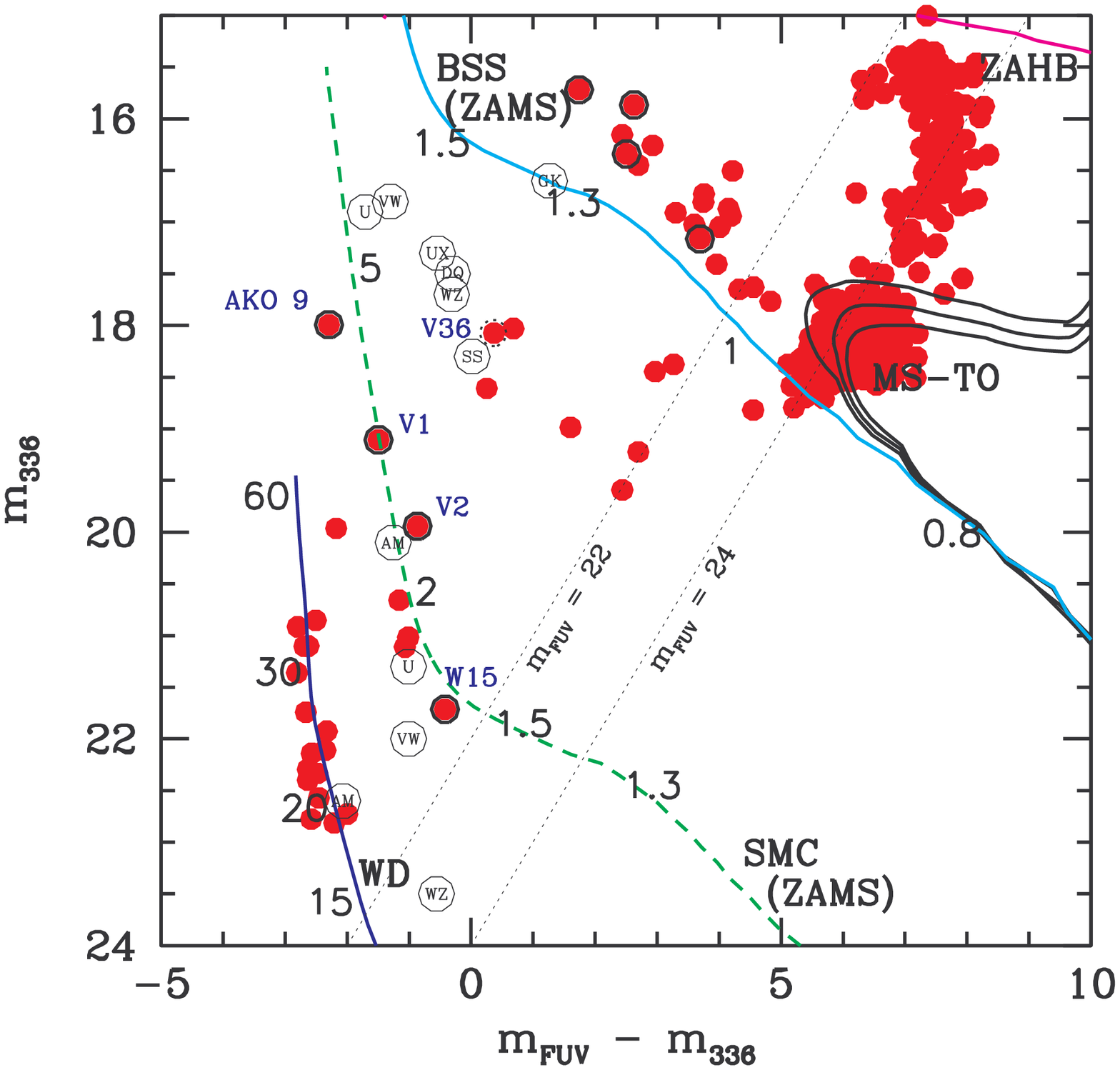,width=4.45in,angle=0}}
\vspace*{8pt}
\caption{The FUV/optical color-magnitude diagram. The $m_{336}$
magnitude corresponds roughly to a U-band magnitude (but measured in
the STMAG system); the $m_{FUV}$ magnitude is derived from the FUV
observations described in the text and is also on the STMAG system
(see Ref~\protect\refcite{knigge1} for details). 
The positions of FUV sources with optical counterparts are
shown as red dots. Variable FUV sources are additionally marked with
black circles, and the four previously known or suspected CVs in our
FoV -- AKO~9, V1, V2 and W15 -- are labelled. The source labelled V36
is another CV candidate and might be the counterpart of a Chandra
X-ray source (see Refs.~\protect\refcite{edmonds1,edmonds2,knigge2} for details). The two short-dashed
diagonal lines are lines of constant FUV magnitude; one ($m_{FUV} =
24$) marks the completeness limit of our catalog, the other ($m_{FUV}
= 22$) is a rough dividing line between WDs, BSs and CVs on the one
hand, and MS turn-off stars, horizontal branch stars and red giants,
on the other. The other lines in the diagram indicate the expected
locations of various stellar populations: WDs, BSs, MS stars (MS-TO
marks the MS turn-off), horizontal branch stars (ZAHB) and SMC stars
(the SMC is located behind 47 Tuc). The numbers next to the
BSs and SMC tracks indicate the masses of stars at the corresponding
location on these tracks; the numbers next to the WD track indicate
the WD temperature at the corresponding location on the track.
Finally, the letters enclosed by open circles mark the positions of
several well-known field CVs if they were observed at the distance and
reddening of 47~Tuc: WZ~=~WZ~Sge; U~=~U~Gem; SS~=~SS~Cyg; VW~=~VW~Hyi;
UX~=~UX~UMa; GK~=~GK~Per; AM = AM~Her; DQ~=~DQ~Her.
Figure reproduced from
Ref.~\protect\refcite{knigge1} (\protect\copyright~2002 The American Astronomical
Society.)
\label{fig3}}
\end{figure}

First, the BS sequence can be seen as a trail of stars starting at the
MS turn-off and moving upwards and to the left. The slight discrepancy
between the observed BS sequence and the synthetic one is to be
expected, since the latter assumes that the BSs lie on the ZAMS. In
reality, BSs are likely to be somewhat evolved, explaining their
location above and to the red of the zero-age main sequence. In total, there are 19 BSs
in our color-magnitude diagram, extending all the way from the MS
turn-off (at approximately 0.9~M$_{\odot}$) to perhaps 1.5~M$_{\odot}$. Four of 
these (circled in Figure~2) appear to be variable in our FUV
photometry. This is reasonable, given that variability among BSs in
47~Tuc has been observed previously.\cite{gilliland} Thus FUV imaging is
indeed an excellent way to locate BSs in dense GC cores. 

The second type of dynamically-formed population present in the CMD is
cataclysmic variables. Given that CVs are binary systems containing an
accreting WD and (usually) a MS star, we expect them to be located
between the WD 
cooling sequence on one side and the MS (and its BS extension) on the
other. We refer to this area of the CMD as the ``CV zone''. There
are 16 objects in the CV zone, not including the star just off the
lower left of the MSTO. How many of these sources are likely to be
real CVs?

Four objects in the CV zone -- AKO~9, V1, V2 and W15 -- are already
strong or confirmed CV candidates. All of these sources are FUV
bright and variable, and all are Chandra 
X-ray sources that were also classified as likely CVs in
Ref.~\refcite{grindlay1}. As discussed further below, AKO~9, V1 and V2
have also already been confirmed spectroscopically as CVs, via the
detection of FUV emission lines (the spectroscopy of AKO~9 has already
been published in Ref.~\refcite{knigge2}). Since we have been able to
easily recover all four previously known CV candidates in our field of
view -- including one that was only recently discovered by Chandra --
we conclude that FUV imaging is indeed a powerful way of finding these
systems. 

The status of the remaining 12 sources in the CV zone is discussed
further in Ref.~\refcite{knigge1}, but the bottom line is that their
nature is still unclear. They probably include some chance
superpositions, some non-interacting 
WD+MS binaries and perhaps one or two background stars located in the Small
Magellanic Cloud behind 47 Tuc. However, some of them might yet turn
out to be CVs, as might additional FUV sources on or near the WD cooling
sequence and FUV sources without any optical counterparts. Depending on
how many of these will ultimately turn out to be CVs, the observed and
predicted (by  tidal capture theory) number of CVs differ by at most a
factor of three, but might even be consistent with each
other.\cite{knigge1}. Recent X-ray based searches for CVs also find
evidence for sizeable populations of CVs in GCs.\cite{grindlay1,grindlay2,pooley1,pooley2}

\subsection{FUV Slitless Spectroscopy: Confirming CV Candidates in 47~Tucanae} 
\label{state_spec}

As already noted above, the great advantage of moving to the FUV
waveband is that the vast majority of ``ordinary'' cluster members are
too cool to show up, thus eliminating the crowding problem that
usually plagues optical surveys of GC cores.

Since crowding is not a serious issue in the FUV, it is possible to
carry out slitless, multi-object spectroscopy of an entire cluster
core and thus obtain spectra for all FUV bright sources
simultaneously. We used this method on 47~Tuc, and Figure~3 shows an
example of the slitless spectra we obtained.

\begin{figure}[th]
\centerline{\psfig{file=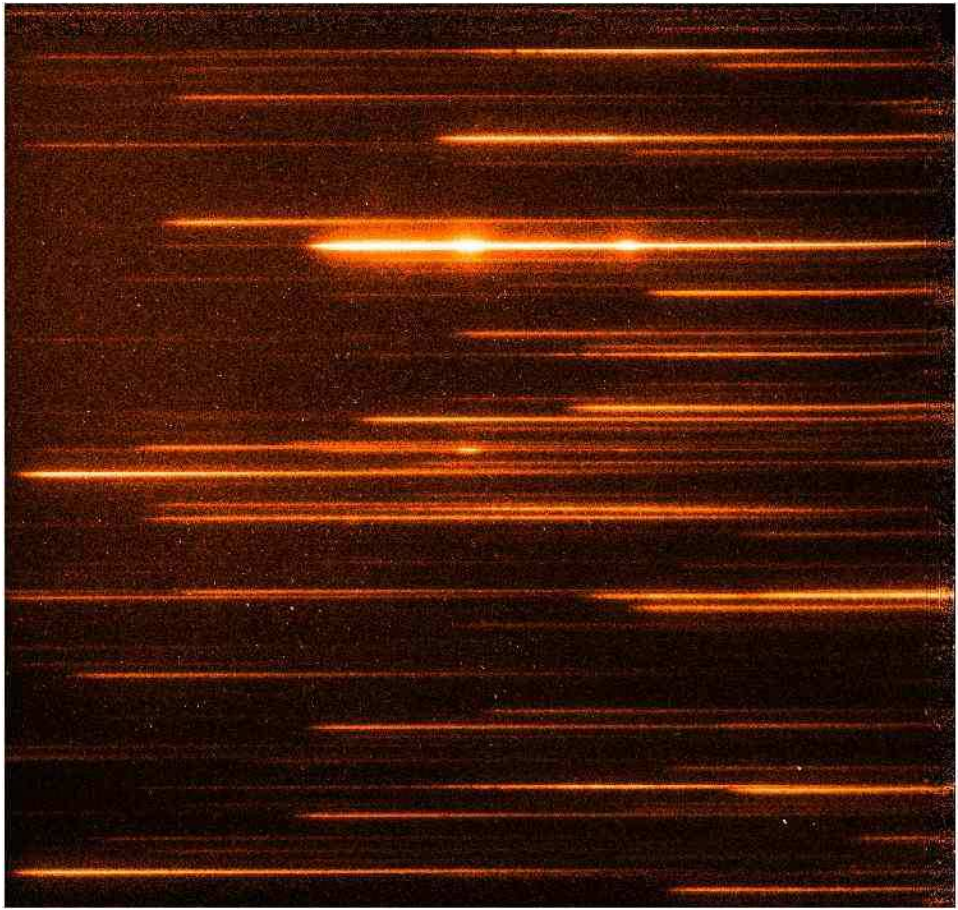,width=5.0in,angle=0}}
\vspace*{8pt}
\caption{The summed 2-D FUV spectral image of 47~Tuc. Each trail in
this figure corresponds the dispersed image of a FUV point source. The
sharp cut-off at the left hand side of each trail is due to the abrupt
decrease in sensitivity around 1450~\AA, where the quartz filter
becomes opaque. The brightest trail is due to the source AKO~9, and
clearly presents two strong emission lines due to C~{\sc iv} and
He~{\sc ii}.}
\end{figure}

Each trail in this image corresponds to the dispersed image of a FUV point
source (c.f. Figure~1). The sharp cut-off at the left hand side of
each trail is due to the abrupt decrease in sensitivity around 1450~\AA,
where the quartz filter becomes opaque. The brightest source in this
spectral image is AKO~9, an object already noted as a likely CV based on
photometric and X-ray data (see previous section). Even a cursory look
at the raw data shows that there are two bright emission lines in its
spectrum, immediately confirming it as a CV. 

Since AKO~9 is extremely bright and relatively isolated, its
spectrum can be extracted fairly straightforwardly from the spectral
images (see Ref.~\refcite{knigge2} for details). Figure~4 shows the
extracted and calibrated FUV spectrum of AKO~9, as
constructed from the FUV spectra obtained in all observing epochs (but
excluding points affected by the eclipse in this high inclination
binary). As expected from the raw, 2-D spectral image in Figure~3, the
calibrated 1-D spectrum contains extremely strong C~{\sc iv} and
He~{\sc ii} emission lines, which are superposed on a blue
continuum. All of these characteristics are consistent with AKO~9
being a dwarf-nova type CV, thus confirming its accreting binary
nature. 

\begin{figure}[th]
\centerline{\psfig{file=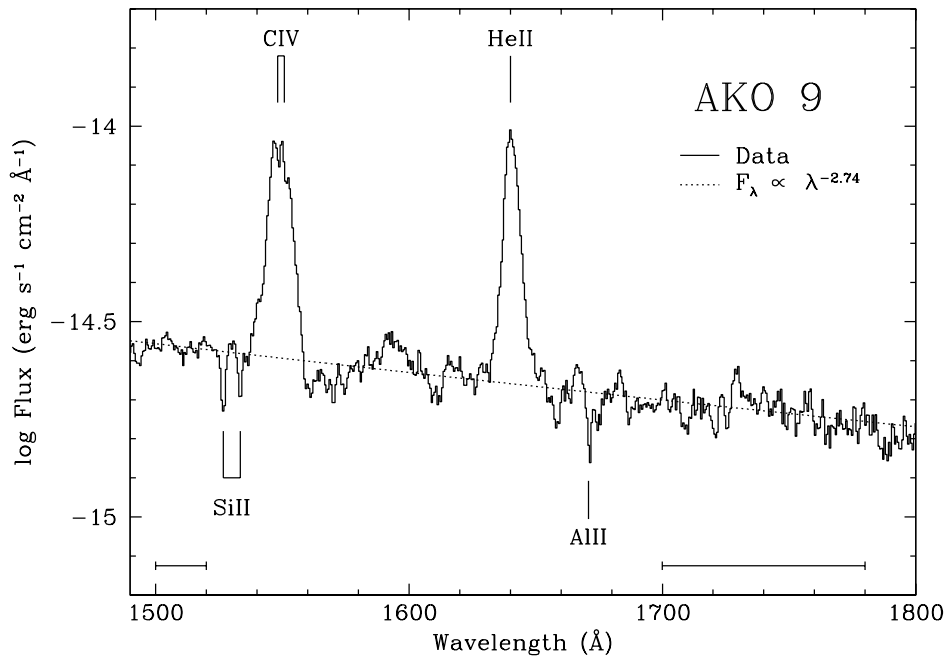,width=3.8in,angle=-90}}
\vspace*{8pt}
\caption{The average out-of-eclipse FUV spectrum of
AKO~9. This spectrum is an exposure-time-weighted average of all FUV
spectra from all observing epochs, excluding only points affected by
the eclipse of this high-inclincation system. The rest wavelengths of
the C~{\sc iv}~doublet 
and He~{\sc ii}~are marked, as are those of the narrow absorption lines
due to Si~{\sc ii}~and Al~{\sc ii}. The Si~{\sc ii}~lines were used to
wavelength calibrate the spectrum. The Al~{\sc ii} line is probably
interstellar. The dotted line shows a power law fit to the continuum
windows indicated by the horizontal bars near the bottom of the
plot. The observed spectrum was corrected for reddening before
carrying out the fit, but the spectra shown here are the uncorrected
data and the reddened power law model. The best-fitting power law
index was $\alpha=-2.74 \pm 0.11$. Thus AKO's FUV spectrum is
extremely blue and presents strong emission lines, confirming it as a
relatively luminous CV. Figure reproduced from Ref.~\protect\refcite{knigge2}
(\protect\copyright~2003 The American Astronomical Society.)}
\end{figure}

Our time-resolved photometry and spectroscopy also allowed us to
finally resolve an almost decade-old puzzle surrounding AKO~9. More
specifically, the authors of Ref.~\refcite{minnitti} observed an "unusual brightening"
of the system, during which its U-band flux increased by 2~mag in less
than two hours. If this brightening is interpreted as a genuine
increase in the system's luminosity, the short time scale of the event
poses serious problems for essentially all models for this system --
including the CV model. However, as described in Ref.~\refcite{knigge2},
our time-resolved data allows us to define a
precise eclipse ephemeris for this system, which in turn shows that
the brightening seen in Ref.~\refcite{minnitti} was, in fact, an eclipse
egress that happened to be observed at a time when AKO~9 was already
in a high state (i.e. during a dwarf nova eruption). The ``unusual
brightening'' thus turns out to be completely consistent with the CV
nature of AKO~9. 

A close look at the spectral image in Figure~4 also reveals another
emission line source, located roughly 20\% below AKO~9. This source
is the dwarf nova-type CV V2.\cite{paresce,shara2} Furthermore, we
have also already confirmed that another previously suspected CV
candidate -- an object known as V1 (see Ref.~\refcite{paresce2}) --
exhibits emission lines (this is not immediately apparent in the 2-D
spectral image). Figure~5 shows quick-and-dirty extractions of the FUV
spectra of these two objects; the emission lines are clearly visible and
confirm the CV status of V1 and V2. The remaining X-ray selected CV
candidate in our imaging field of view (denoted W15 in
Ref.~\refcite{grindlay1}) is unfortunately just outside the
spectroscopic field of view.

\begin{figure}[th]
\centerline{\psfig{file=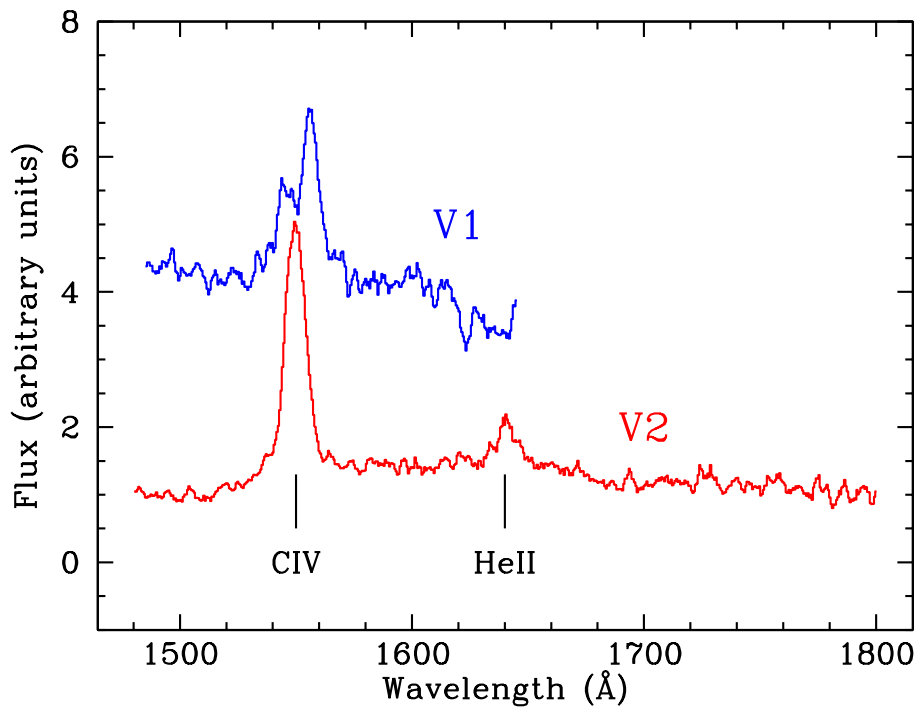,width=3.8in,angle=-90}}
\vspace*{8pt}
\caption{The FUV spectra of the CV candidates V1 and V2 in 
47~Tuc. Even though these are quick-and-dirty spectral extractions
(without absolute flux calibration), it is clear that both sources
display the strong FUV emission lines that are a hallmark of CVs.}
\end{figure}

The example of AKO~9 (and also V1 and V2) nicely demonstrates the
power of moving to the FUV: by shifting to this waveband, we were able
to obtain time-resolved, multi-object spectroscopy simultaneous for
AKO~9, V1, V2 and all other bright FUV sources. This powerful and
efficient technique would be impossible to apply in the optical
region, due to the severe crowding there. 

\section{The Future}

\subsection{FUV Imaging}
\label{future_image}

Our FUV imaging observations of 47~Tuc clearly illustrate the unique
advantages offered by this waveband for finding and studying the
dynamically-formed stellar populations in GCs. We have therefore
already been granted additional HST time to observe five other GCs
with HST/STIS and/or HST/ACS. In addition, two GCs -- NGC~2808 and
NGC~6881 -- are actually used as calibration targets by HST and thus
already have extensive FUV imaging data sets in the HST archive. Thus
we are well on the way towards the construction of a sizeable archive
of FUV imaging data on GCs.

Most of this data is already in hand, and our group is working hard to
reduce and analyse it all. As one example of the progress currently
being made, Figure~6 shows the combined FUV image of NGC~2808. As
already noted above, this data set has already been analyzed in
Ref.~\refcite{brown} with an emphasis on hot subluminous horizontal branch
stars. However, their FUV/NUV color-magnitude diagram clearly showed
both a BS sequence and objects in the CV zone, so we are now
revisiting these observations with a focus on the dynamically-formed
stellar populations (e.g. by means of searching for variability among 
the BSs and CV candidates; Dieball et al., in preparation). Similar
progress is being made on the other clusters for which FUV data is
already available, and initial FUV-imaging publications should soon be
available on M15 (Shara et al., submitted [confirmation of one
particular CV]; Dieball et al., in preparation [work on the full FUV
data set]), NGC~1851 and NGC~6681 (Zurek et al., in preparation), and
also NGC~6752 and NGC~6397 (Shara et al., in preparation). Thus the
next year or so should see rapid growth in this exciting area.

\begin{figure}[th]
\centerline{\psfig{file=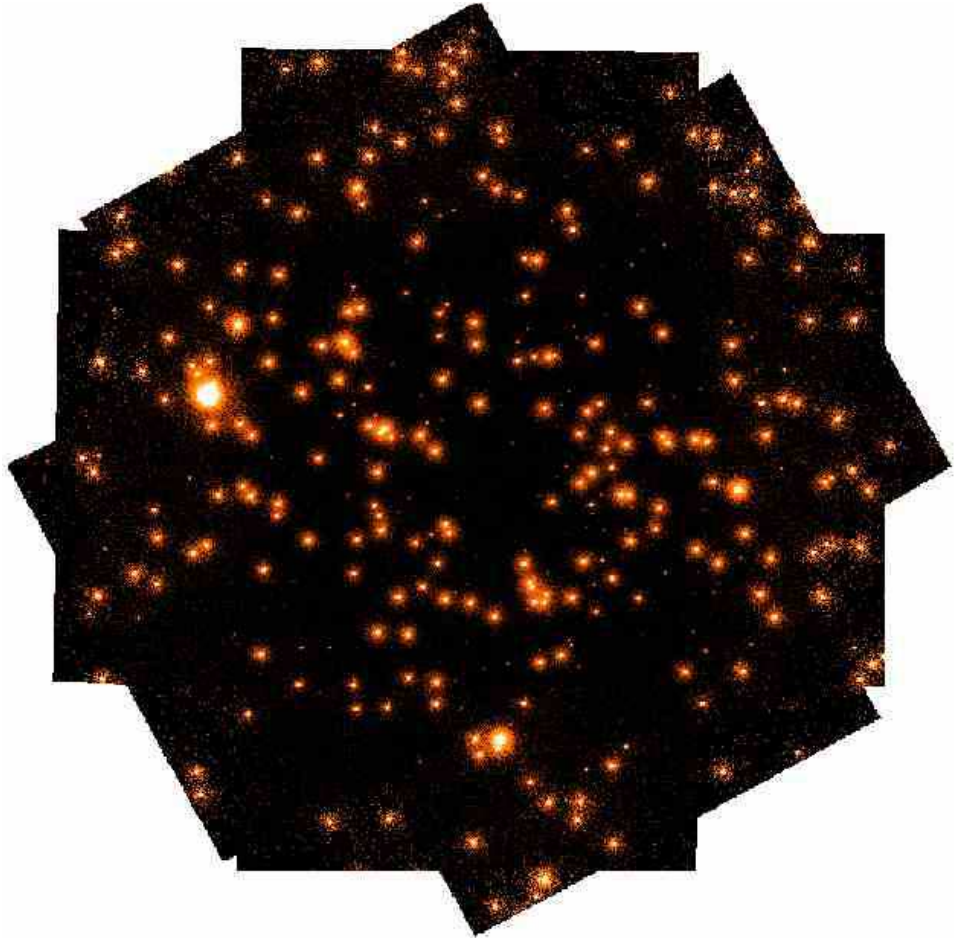,width=5.0in}}
\vspace*{8pt}
\caption{The combined FUV image of the globular cluster NGC~2808. The
strange shape of this image reflects the fact that it is a mosaic
constructed from several exposures that were rotated with respect to
each other. The brightest sources in the image are blue horizontal
branch stars, whose properties have been explored in
Ref.~\protect\refcite{brown}. However, the color-magnitude diagram presented
in Ref.~\protect\refcite{brown} also reveals several BSs, as well as objects
between the WD and main sequences that might be CVs. Our ongoing
reanalysis of this data set thus focuses particularly on these
dynamically-formed stellar populations.} 
\end{figure}

\subsection{FUV Slitless Spectroscopy}

As described in Section~\ref{state_spec}, the lack of crowding in the
FUV waveband makes it possible to carry out slitless, multi-object
spectroscopy even in the core of a GC. However, the results we have so
far obtained from our spectroscopic survey of 47~Tuc -- in particular,
confirmation of AKO~9, V1 and V2 as CVs by virtue of strong emission
lines in their FUV spectra  -- still only scratch the surface of this
phenomenal data set. Most importantly, our spectral images contain
reasonable quality spectra of many additional objects in the ``CV
zone'' (see Section~\ref{state_image}) and thus might allow us to
confirm or rule out the CV nature of these objects. In addition, many
BSs and WDs should be bright enough to allow their spectra to be
extracted. However, these sources are generally fainter and less
well isolated than AKO~9, so extracting their spectra is considerably
more difficult. In particular, the extraction must take into
account blending between neighboring sources. Luckily, spectral
extraction software for this type of data already
exists.\cite{miskey}. We are therefore currently working with the
authors of the software to apply their technique to our data. Thus FUV
spectra of several additional CV candidates and BSs in 47~Tuc will
soon become available (Knigge et al., in preparation).

\section{Conclusions}

The goal of this short review has been to demonstrate the great
benefits of moving to the FUV waveband when studying hot, 
dynamically-formed stellar populations in globular clusters. The most
important advantage of the FUV waveband is that ordinary cluster
members are too cool to show up as FUV sources, whereas all of the
most interesting stellar populations in GCs -- blue stragglers, accreting
binaries, but also young white dwarfs -- are strong FUV emitters. As a
result, FUV images suffer from much less crowding than optical ones,
and almost every moderately bright FUV source is bound to be an 
interesting object. These expected advantages are completely borne
out by our first FUV imaging survey of 47~Tuc, in which we have been
able to identify many BSs and CV candidates. 

Since crowding is not a problem in the FUV, we have even been able to
carry out slitless, multi-object spectroscopy in the very core of
47~Tucanae. This is an extremely efficient way of obtaining
spectroscopic classifications of interesting sources, but is
completely impossible in the optical waveband. In 47~Tuc, our FUV spectral 
images have already allowed us to confirm the CV nature of several
bright FUV sources in this cluster. 

In closing, I would like to emphasize that deep FUV imaging and
spectroscopy of GCs is extremely powerful and is definitely a growth
industry. Thus while deep FUV observations have been published for
only two clusters so far, work by our group alone should at least
triple this number over the coming months. 

The FUV project described in this review is very much a joint effort
with my collaborators Dave Zurek, Mike Shara, Knox Long, Ron Gilliland
and Andrea Dieball. The financial support of the UK's Particle Physics and
Research Council for my work in this area is gratefully
acknowledged. Much of this review has been based on 
Refs.~\refcite{knigge1,knigge2}, which are \copyright~2002,2003 The
American Astronomical Society (all rights reserved).

%\section*{References}

\end{document}